\def\Ic{{\text{sgn}\left (\mathfrak{I_c} \right )}}

\documentclass[aps,prappl,twocolumn,groupedaddress,showkeys,showpacs,superscriptaddress,floatfix,longbibliography]{revtex4-1}
\usepackage{epsfig}
\usepackage{multirow}
\usepackage{amsmath,amssymb,mathtools}
\usepackage{color}
\usepackage{graphicx}
\usepackage{dcolumn}
\usepackage{bm}
\usepackage{subfigure}
\usepackage[utf8]{inputenc}
\usepackage[T1]{fontenc}
\usepackage{tikz}

\graphicspath{{Figures/}}
\definecolor{AB-color}{RGB}{128,0,128}
\begin{document}

\title{Non-linear critical current thermal response of an asymmetric Josephson tunnel junction}

\author{Claudio Guarcello}
\email{claudio.guarcello@nano.cnr.it}
\affiliation{NEST, Istituto Nanoscienze-CNR and Scuola Normale Superiore, Piazza San Silvestro 12, I-56127 Pisa, Italy}
\author{Alessandro Braggio}
\affiliation{NEST, Istituto Nanoscienze-CNR and Scuola Normale Superiore, Piazza San Silvestro 12, I-56127 Pisa, Italy}
\author{Paolo Solinas}
\affiliation{SPIN-CNR, Via Dodecaneso 33, 16146 Genova, Italy}
\author{Francesco Giazotto}
\affiliation{NEST, Istituto Nanoscienze-CNR and Scuola Normale Superiore, Piazza San Silvestro 12, I-56127 Pisa, Italy}

\date{\today}

\begin{abstract}
We theoretically investigate the critical current of a thermally-biased SIS Josephson junction formed by electrodes made by different BCS superconductors. The response of the device is analyzed as a function of the asymmetry parameter, $r=T_{c_1} /T_{c_2}$. We highlight the appearance of jumps in the critical current of an asymmetric junction, namely, when $r\neq1$. In fact, in such case at temperatures at which the BCS superconducting gaps coincide, the critical current suddenly increases or decreases. In particular, we thoroughly discuss the counterintuitively behaviour of the critical current, which increases by enhancing the temperature of one lead, instead of monotonically reducing. In this case, we found that the largest jump of the critical current is obtained for moderate asymmetries, $r\simeq3$. In view of these results, the discussed behavior can be speculatively proposed as a temperature-based threshold single-photon detector with photon-counting capabilities, which operates non-linearly in the non-dissipative channel.
\end{abstract}

\maketitle

\section{Introduction}
\label{Sec00}\vskip-0.2cm

\begin{figure}[b!!]
\centering
\includegraphics[width=0.8\columnwidth]{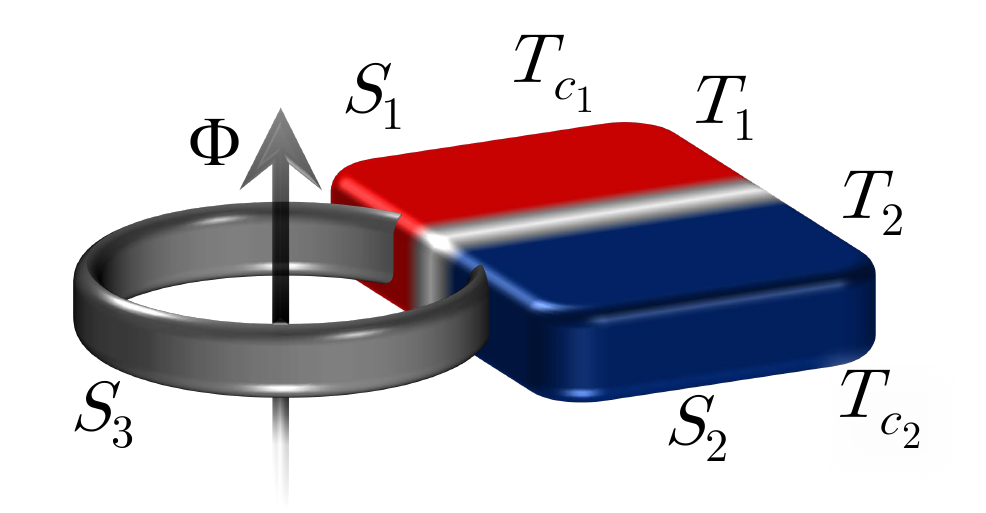}
\caption{Schematic illustration of a temperature-biased SIS Josephson tunnel junction formed by the superconducting leads $S_1$ and $S_2$, with critical temperatures $T_{c_1}$ and $T_{c_2}$, and residing at temperatures $T_1$ and $T_2$. The junction is enclosed in a superconducting ring pierced by a magnetic flux $\Phi$ which allows phase biasing of the weak link. The ring is supposed to be made by a third superconductor $S_3$ with energy gap $\Delta_3\gg\Delta_1,\Delta_2$ so to suppress the heat losses.}
\label{Fig01}
\end{figure}

More than 50 years after its discovery, the Josephson's effect~\cite{Jos61,And63} is still a province able to provide intriguing, even unexpected, physical phenomena, from which novel devices are continuously conceived. This is the case of the plethora of works descending only recently~\cite{Gia12,Mar14,MarSol14,ForGia17} from the earlier intuition that a temperature bias imposed across a Josephson junction (JJ) produces a phase-dependent heat flow through the device~\cite{Mak65}. We are dealing with the \emph{phase-coherent caloritronics}~\cite{Gia06,Mes06,MarSol14,ForGia17}, namely, an emerging research field from which fascinating Josephson-based devices, such as heat interferometers~\cite{GiaMar12,Gia12} and diffractors~\cite{Gia13,Mar14,Gua16}, heat diodes~\cite{Mar15} and transistors~\cite{For16}, solid-state memories~\cite{Gua17,GuaSol17,GuaSol18}, microwave refrigerators~\cite{Sol16}, thermal engines~\cite{Pao18}, thermal routers~\cite{Tim18,Gua18}, heat amplifier~\cite{Pao17}, and heat oscillator~\cite{GuaSolBra18}, were recently designed and actualized. Even the critical current $I_c$ of a Josephson tunnel junction, namely, the maximum dissipationless current that can flow through the device, deviates from the well-known Ambegaokar-Baratoff relation~\cite{Amb63} in the presence of a thermal bias imposed across the junction, namely, as the superconducting electrodes reside at different temperatures, as portrayed in Fig.~\ref{Fig01}.

In this work we explore peculiar features of the critical current of a thermally-biased asymmetric tunnel JJ. We theoretically demonstrate that the critical current $I_c$ of a junction formed by different superconductors shows steeplike variation and it is asymmetric in the temperature switch. Specifically, we show that the critical current suddenly jumps at specific temperatures at which the BCS superconducting gaps~\cite{Bar82,Lik86} are equal. The abrupt variations are due to the matching in the singularities of the anomalous Green functions in the two superconductors~\cite{Bar82}. This feature is the non-dissipative counterpart of the discontinuities discussed in the quasiparticle current flowing through a voltage-biased $\text{S}_1\text{I}\text{S}_2$ junction~\cite{Bar82,Har74} and the heat current flowing through a temperature biased junction~\cite{Gol13,For16}, both stemming from the alignment of the singularities of the BCS DOSs in the superconductors~\cite{Bar82}.

We observe that sudden decreases in the critical current were already noted, but not extensively discussed so far~\cite{Gia15,For16}. Additionally, for appropriate parameters values we will show that the critical current counterintuitively behaves, since it increases by enhancing the temperature, instead of decreasing. Furthermore, we study the asymmetry of the critical current with respect to the switching of the temperatures, through the definition of a suitable temperature-switching asymmetry parameter.
We also discuss the linear regime in response to a thermal gradient, by studying the first-order coefficients of the critical current expansion as a function of the average temperature, at a few values of the Dynes parameter.

Finally, according to the step-like behavior of the critical current, we suggest the application of this device as a non-dissipative threshold single-photon detector, based on the sudden increase of $I_c$ due to a photon-induced heating of one of the electrodes of the junction.

The paper is organized as follows. In Sec.~\ref{Sec01}, we study the behavior of the critical current by varying non-linearly the temperatures of the device and the ratio between the critical temperatures of the two superconductors. In Sec.~\ref{Sec02}, we address the linear approximation in the temperature gradient. We discuss in Sec.~\ref{Sec03} a possible applications of the discussed effects as a single-photon detector. 
In Sec.~\ref{Sec04}, the conclusions are drawn.

\section{The critical current}
\label{Sec01}\vskip-0.2cm

Here, we explore how the critical current of a temperature-biased SIS JJ depends on the superconductors composing the device. Indeed, we consider a junction formed by different BCS superconductors, so that we can define an asymmetry parameter 
\begin{equation}
r=\frac{T_{c_1}}{T_{c_2}}=\frac{\Delta_{10}}{\Delta_{20}},
\end{equation}
where $T_{c_j}$ is the critical temperature and $\Delta_{j0}=1.764 k_BT_{c_j}$ is the zero-temperature superconducting BCS gap~\cite{Tin04} of the $j$-th superconductor (with $k_B$ being the Boltzmann constant).

A Josephson tunnel junction formed by two superconducting leads $S_1$ and $S_2$ with energy gaps $\Delta_1$ and $\Delta_2$ residing at temperatures $T_1$ and $T_2$, see Fig.~\ref{Fig01}, can support a non-dissipative Josephson current~\cite{Bar82}
\begin{equation}\label{JosephsonCur}
I_\varphi\left ( T_1,T_2 \right )=I_c\left ( T_1,T_2 \right )\sin\varphi,
\end{equation}
with $\varphi$ being the macroscopic quantum phase difference between the superconductors across the junction, and $I_c( T_1,T_2)$ being the critical current, which reads~\cite{Gol04,Gia05,Tir08,Bos16}
\begin{eqnarray}\label{IcT1T2Im}\nonumber
I_c\left ( T_1,T_2 \right )=&&\frac{1}{2eR}\Bigg |\underset{-\infty}{\overset{\infty}{\mathop \int }} \Big \{ f\left ( \varepsilon ,T_1 \right )\textup{Re}\left [\mathfrak{F}_1(\varepsilon,T_1 ) \right ]\textup{Im}\left [\mathfrak{F}_2(\varepsilon,T_2 ) \right ] \\&&+ f\left ( \varepsilon ,T_2 \right )\textup{Re}\left [\mathfrak{F}_2(\varepsilon,T_2 ) \right ]\textup{Im}\left [\mathfrak{F}_1(\varepsilon,T_1 ) \right ]\Big \} d\varepsilon\Bigg |.
\end{eqnarray}

Here, $R$ is the normal-state resistance of the junction, $e$ is the electron charge, $f\left ( \varepsilon ,T_j \right )=\tanh\left ( \varepsilon/2 k_B T_j \right )$, and
\begin{equation}\label{Green}
\mathfrak{F}_j(\varepsilon,T_j ) =\frac{\Delta_j \left ( T_j \right )}{\sqrt{\left ( \varepsilon +i\Gamma_j \right )^2-\Delta_j^2 \left ( T_j\right )}}
\end{equation}
is the anomalous Green's function of the $j$-th superconductor~\cite{Bar82}, with $\Gamma_j=\gamma_j\Delta_{j0}$ being the Dynes parameter~\cite{Dyn78}. The so-called Dynes model~\cite{Dyn78,Dyn84} is based on an expression
of the BCS DOS including a lifetime broadening. It allows to take into account the smearing of the \emph{IV} characteristics of a JJ, that is the persistence of a small subgap current at low voltages. In fact, a nonvanishing $\gamma_j$ introduces effectively states within the gap region, $\left | \varepsilon \right |<\Delta_j$, as opposed to the ideal BCS DOS obtained at $\gamma_j=0$, which instead results in a vanishing DOS within the gap~\cite{Pek10,Sai12}. Unless otherwise stated, hereafter we assume $\gamma_1 =\gamma_2 =\gamma =10^{-4}$, namely, a value often used to describe realistic superconducting tunnel junctions~\cite{Pek04,Mar15,For16}.

\begin{figure*}[t]
\centering
\includegraphics[width=\textwidth]{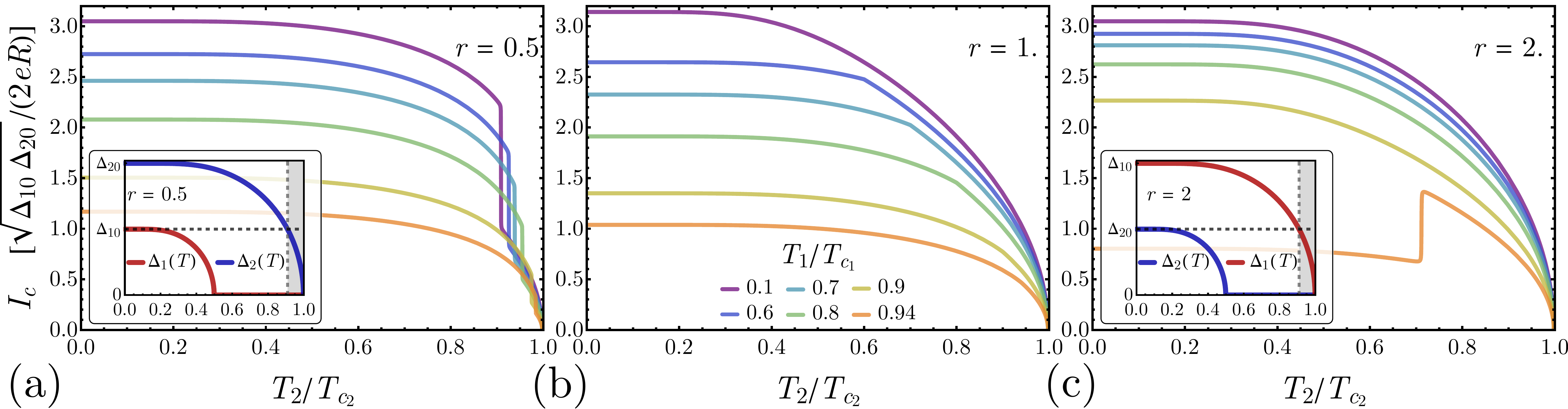}
\caption{Critical current, in units of $\sqrt{\Delta_{10}\Delta_{20}}/(2eR)$, as a function of the normalized temperature $T_2/T_{c_2}$ at a few values of the normalized temperature $T_1/T_{c_1}$, for $r=0.5$, $1$, and $2$, see panel (a), (b), and (c), respectively. Insets in panels (a) and (c) show the superconducting gaps $\Delta_{1}$ and $\Delta_{2}$ as a function of the temperature $T$, normalized to $\text{Max}\{T_{c_1},T_{c_2}\}$, for $r=0.5$ and $2$, respectively. Legend in panel (b) refers to all panels.}
\label{Fig02}
\end{figure*}

Fig.~\ref{Fig01} shows a possible experimental realization of the discussed setup where we clearly indicate how to master the phase difference across the device. The thermally-biased junction is enclosed, through clean contacts, within a superconducting ring pierced by a control magnetic flux $\Phi$. In this way, we achieve the phase-biasing via this external flux, which allows us to thoroughly play with the macroscopic phase difference across the JJ. In fact, neglecting the ring inductance, the phase-flux relation is given by $\varphi=2\pi\Phi/\Phi_0$~\cite{Cla04}($\Phi_0= h/2e\simeq2\times10^{-15}\; \textup{Wb}$ is the magnetic flux quantum, with $h$ being the Planck constant). Accordingly, the phase drop across the junction can vary within the whole phase space, i.e., $-\pi\leq\varphi\leq\pi$. The ring is supposed to be made by a third superconductor $S_3$ with energy gap $\Delta_3\gg\Delta_1,\Delta_2$ so to suppress the heat losses thanks to Andreev reflection heat mirroring effect~\cite{And64}.

\subsection{Non-linear temperature behavior of $I_c$}
We first study the critical current of the device by choosing the asymmetry parameter $r$, and changing the temperature of $S_2$ at fixed $T_1$ for non-linear regimes of temperatures. 

The behavior of the critical current $I_c$, in units of $\sqrt{\Delta_{10}\Delta_{20}}/(2eR)$, as a function of the normalized temperature $T_2/T_{c_2}$ at a few values of the normalized temperature $T_1/T_{c_1}$, for $r=\{0.5,1,2\}$ is shown in Fig.~\ref{Fig02}. 
We note that the critical current generally reduces by increasing $T_1$, an effect that may be naively interpreted as the usual detrimental effect of the temperature on the critical current. Anyway, we will see that for $r\neq1$, the temperature may affect the critical current in an unexpected way. In fact, we observe that the critical current as a function of $T_2$ may present a steeplike response. Specifically, for $r\neq1$, i.e., the asymmetric junction case, curves may show jumps, see Figs.~\ref{Fig02}(a) and (c), whereas for a symmetric junction, namely, $r=1$, curves present only a change of slope, see Fig.~\ref{Fig02}(b). These distinctive behaviors stem from the alignment of the singularities in the Green's functions $\mathfrak{F}_j$ at $\varepsilon=\Delta_j$ when 
\begin{equation}\label{singularities}
\Delta_1(T_1)=\Delta_2(T_2).
\end{equation}

In order to correctly interpret this phenomenology, we discuss first the critical currents for $r<1$, i.e., $r=0.5$ shown in Fig.~\ref{Fig02}(a). In this case, the superconducting gap $\Delta_1$ is smaller than $\Delta_2$, namely, $\Delta_1(T)<\Delta_2(T)\; \forall T\in[0-T_{c_2}]$, see the inset of Fig.~\ref{Fig02}(a), so that for each temperature $T_1$ certainly exists a temperature $T_2$ satisfying Eq.~\eqref{singularities}. However, this condition is fulfilled only when $T_2$ is higher then the threshold value $T_2^{\text{th}}$ at which $\Delta_2(T_2^{\text{th}})=\Delta_{10}$ (where $\Delta_{10}=\Delta_1(T_1=0)$). Specifically, for $r=0.5$, one obtains $T_2^{\text{th}}\simeq0.91\;T_{c_2}$, see dashed lines in the inset of Fig.~\ref{Fig02}(a). Therefore, the sharp jumps in the critical current emerge at $T_2> T_2^{\text{th}}$, namely, at the $T_2$'s values within the shaded region in the inset of Fig.~\ref{Fig02}(a). We note that the height of the jumps reduces by increasing $T_1$~\footnote{We note that the sharpness of the jump depends on the value of the Dynes's parameters, as we will discuss in detail later.}.

In the symmetric case, namely, $r=1$, shown Fig.~\ref{Fig02}(b), the condition~\eqref{singularities} can be satisfied only at $T_1=T_2$. In this case, there is no jump, so that the curves have a change of slope, in the place of a jump, at $T_1=T_2$. 

Finally, for $r>1$, i.e., $r=2$ in Fig.~\ref{Fig02}(c), $\Delta_1(T)>\Delta_2(T)\; \forall T\in[0-T_{c_1}]$, see the inset of Fig.~\ref{Fig02}(c), so that the condition~\eqref{singularities} is fulfilled only at temperatures $T_1$ higher than the value $T_1^{\text{th}}$ at which $\Delta_1(T_1^{\text{th}})=\Delta_{20}$ (where $\Delta_{20}=\Delta_2(T_2=0)$), see the shaded region in the inset of Fig.~\ref{Fig02}(c). Specifically, for $r=2$, one obtains $T_1^{\text{th}}\simeq0.91\;T_{c_1}$. Indeed, among those shown in Fig.~\ref{Fig02}(c), only the curve at $T_1=0.94\;T_{c_1}$ shows a jump. Interestingly, in this case the critical current $I_c$ behaves counterintuitively, since by raising the temperature it sharply increases undergoing a jump, instead of decreasing monotonically. Moreover, this positive jump becomes higher at a temperature $T_1$ just above $T_1^{\text{th}}$ and reduces by further increasing it. This odd behaviour of the critical current can be anticipated also by further inspecting Fig.~\ref{Fig02}(a), since the point where the jump is located, i.e. $T_2^J$, shifts towards higher temperatures by increasing $T_1$. So, by inverting the role of $T_1$ and $T_2$ the jumps showed in Fig.~\ref{Fig02}(a) would necessary imply the behaviour shown in Fig.~\ref{Fig02}(c).

We note that, both in $r>1$ and $r<1$ cases, the temperature ranges in which the jumps in $I_c$ appear can be enlarged by reducing the temperatures $T_1^{\text{th}}$ and $T_2^{\text{th}}$, namely, by considering junctions less and less asymmetric, i.e., $r\to1$. Nonetheless, in this case the heigh of the jumps tends to reduce, up to vanish just for $r=1$. Conversely, by increasing the asymmetry between the gaps, namely, for $r\gg1$ (or $r\ll1$), we are suppressing one superconducting gap with respect to the other. In these cases, $T_1^{\text{th}}\to T_{c_1}$ (or $T_2^{\text{th}}\to T_{c_2}$), and the ranges of temperature in which the $I_c$ jumps appear get narrower. Accordingly, since $I_c\to0$, we expect that, also in these regimes, the height of the $I_c$ jumps will tend to diminish.

\begin{figure}[t!!]
\centering
\includegraphics[width=\columnwidth]{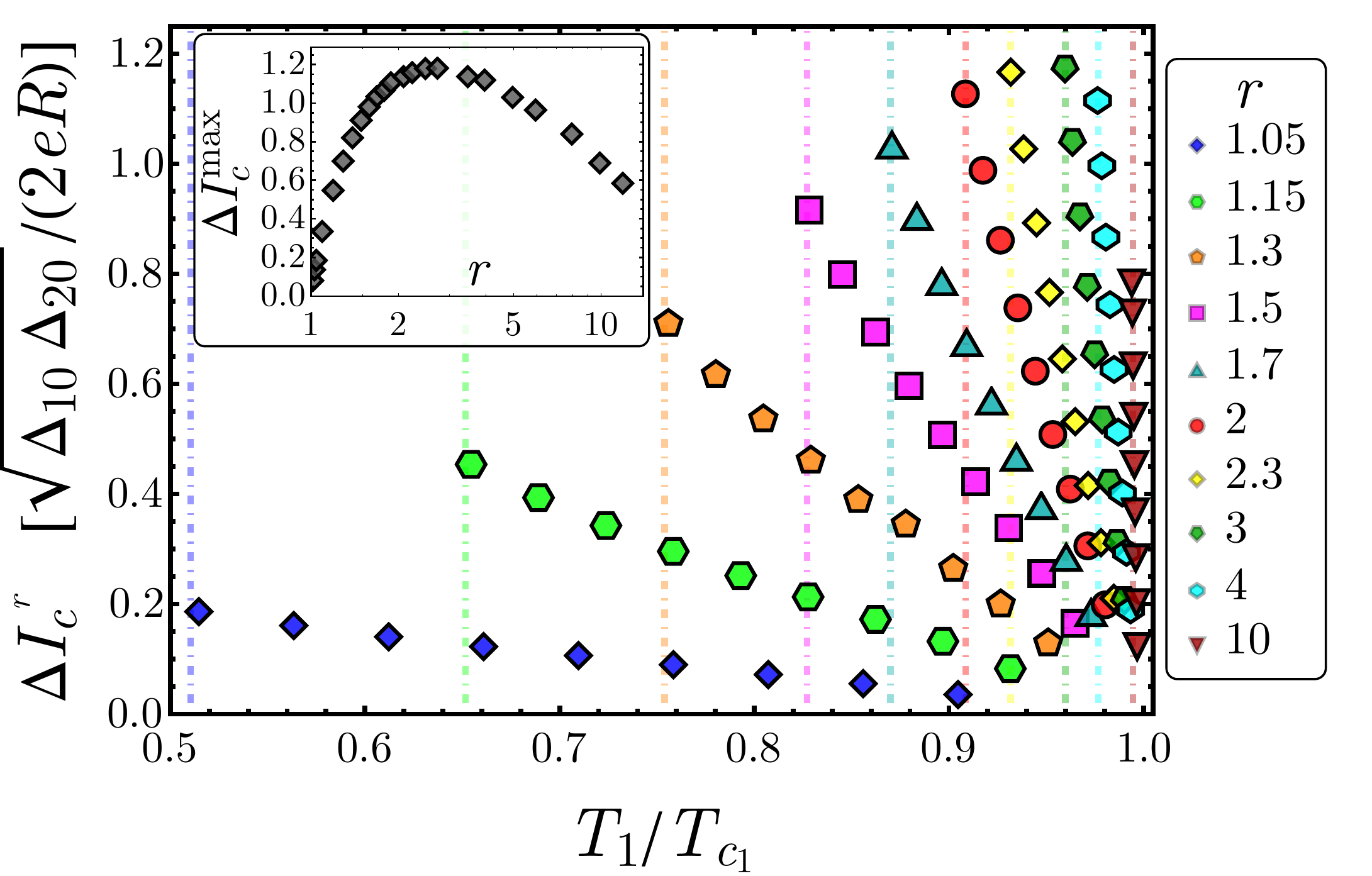}
\caption{Height of the critical current jump, see Eq.~\eqref{DeltaIc}, in units of $\sqrt{\Delta_{10}\Delta_{20}}/(2eR)$, as a function of $T_1/T_{c_1}$ at a few values of $r\in]1\div10]$. The vertical dotted lines indicate the threshold temperatures $T_1^{\text{th}}/T_{c_1}$ above which the jumps of $I_c$ appear, calculated at the values of $r$ used in the figure. In the inset: maximum value of $\Delta I_c^{r}$, i.e., $\Delta I_c^{\text{max}}=\text{max}_{\,T_1}\Delta I^{r}_c(T_1)$, in units of $\sqrt{\Delta_{10}\Delta_{20}}/(2eR)$, as a function of $r$.}
\label{Fig03}
\end{figure}

In light of this, we investigate the dependence of the height of the critical current jump, $\Delta I^{r}_c(T_1)$, on the asymmetry parameter $r$ by varying the temperature $T_1$. Specifically, we explore the cases for $r>1$, namely, the cases giving positive jumps of $I_c$, as already discussed in Fig.\ref{Fig02}(c). In fact, for $r>1$, at a temperature $T_2=T_2^J$ satisfying Eq.~\eqref{singularities}, the critical current $I_c(T_1>T_1^{\text{th}},T_2^J)$ sudden increases. In this case, we additionally observe that $I_c$ has a minimum just before the jump, i.e., for $T_2<T_2^J$, and a maximum just after the jump, i.e., for $T_2>T_2^J$. Therefore, we define the height of the critical current jump as the difference between these maximum and minimum $I_c$ values, namely,
\begin{equation}
\label{DeltaIc}
\Delta I^{r}_c(T_1)\text{=}\text{max}_{\,T_2}I_{c}(T_1,T_2>T^J_2)-\text{min}_{\,T_2}I_{c}(T_1,T_2<T^J_2),
\end{equation}
where $T^J_2$ is the temperature $T_2$ at which the jump occurs, $T_1>T_1^{\text{th}}$, and $r>1$. The behavior of $\Delta I^{r}_c(T_1)$ as a function of $T_1$ at a few values of $r$ is shown in Fig.~\ref{Fig03}. The vertical dashed-dotted lines indicate the threshold temperatures $T_1^{\text{th}}/T_{c_1}$ above which the jumps of $I_c$ appear, calculated at the values of $r$ used in the figure. We observe that, at a given $r$, $\Delta I^{r}_c(T_1)$ is maximal for a $T_1$ just above $T_1^{\text{th}}$ and than it reduced linearly by increasing $T_1$ up to vanishes for $T_1\to T_{c_1}$. Interestingly, we observe that the maximum value of $\Delta I^{r}_c(T_1)$, calculated as $\Delta I_c^{\text{max}}=\text{max}_{\,T_1}\Delta I^{r}_c(T_1)$, behaves non-monotonically by increasing $r>1$, approaching zero for $r\to1$ and $r\gg1$ and reaching a maximum for $r\simeq3$, as shown in the inset of Fig.~\ref{Fig03}. Accordingly, the highest $I_c$ jump is obtained for $T_{c_1}\simeq3T_{c_2}$. 

To have an idea of the situation in which the present effect can be observed, we assume a JJ with a barrier resistance of $R=100\; \Omega$ between, for instance, Nb ($T_{c_1}=9.2\;\text{K}$) and Ta ($T_{c_2}=4.4 \;\text{K}$), corresponding to an asymmetry parameter of $r\approx 2$, one finds a jump of $\Delta I_c^{r=2}\simeq 1.2\sqrt{\Delta_{10}\Delta_{20}}/(2eR)\simeq 5.8\;\mu\text{A}$, when the maximal critical current at low temperatures is $I_{c,\text{max}}^{r=2}\simeq 3.05\sqrt{\Delta_{10}\Delta_{20}}/(2eR)=14.5\;\mu\text{A}$, see Fig.~\ref{Fig02}(c).
Nonetheless, we observe that in this case the range of $T_1$ at which the jump of $I_c$ emerges is very nearby to the critical temperature.

In the previous discussion we analyzed the jump for $r>1$, although one can easily generalize the previous results also to the $r<1$ case, due to the discussed symmetry between the $r<1$ and $r>1$ cases by exchanging the role of the temperatures $T_1$ and $T_2$. In particular, for $r<1$ the jump height will be maximum for $r\simeq1/3$. In this case, the value of $T_2$ at which the jump appears is really nearby the critical temperature $T_{c_2}$, as can be easily seen by comparing Fig.~\ref{Fig02}(a) with Fig.~\ref{Fig02}(c). 

\begin{figure}[t!!]\label{Ic_ManyGamma}
\centering
\includegraphics[width=0.89\columnwidth]{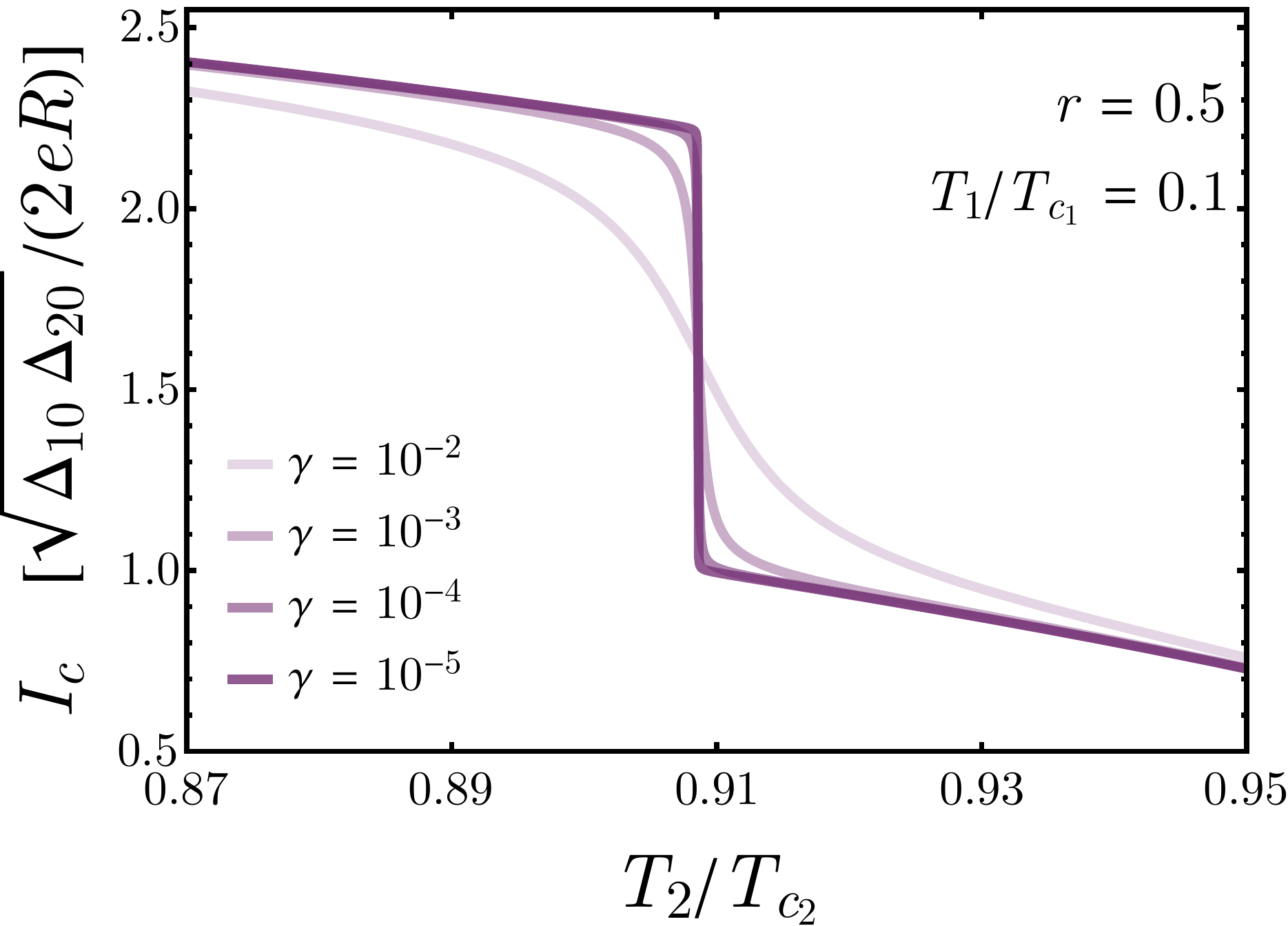}
\caption{Critical current, in units of $\sqrt{\Delta_{10}\Delta_{20}}/(2eR)$, as a function of $T_2/T_{c_2}$ at a few values of the Dynes parameter $\gamma$, for $r=0.5$ and $T_1/T_{c_1}=0.1$.}
\label{Fig04}
\end{figure}

The impact of the Dynes parameter, $\gamma$, on the critical current is highlighted in Fig.\ref{Fig04}. In this figure, the behavior of $I_c$, in units of $\sqrt{\Delta_{10}\Delta_{20}}/(2eR)$, as a function of $T_2/T_{c_2}$ at a few values of $\gamma$, for $r=0.5$ and $T_1/T_{c_1}=0.1$, is shown. Specifically, we evidence how the critical current changes by varying $\gamma$ in a neighborhood of a jump. We observe that the higher the $\gamma$ value, the smoother the $I_c$.~\footnote{Since the detection readout can be done in the non-dissipative regime, as we will discuss later, any $I_c$ change not affects the thermal exchanges.}

In Sec.~\ref{Sec03} we will discuss some possible applications of this device, but certainly the sharpness of the jump is an important figure of merit, which is potentially connected to the sensitivity of the junction to small temperature variations around the operating point $T_2^J$. Higher sensitivities in temperature can be obtained by maximizing the jump sharpness, i.e., by increasing the current jump height $\Delta I_c$ and/or by minimizing the Dynes parameter~\cite{Pek10}. We note that, in the perspective of detecting small variations of $T_2$, it is more convenient to consider the case where the jump, as a function of $T_2$, is positive, as shown in Fig.~\ref{Fig02}(c) for $r>1$, since the normalized temperature $T_2^J/T_{c2}$ is smaller than the case with $r<1$. 
This means to keep the superconducting electrode with the higher $T_c$ at a temperature quite near to the critical value, and to leave free the temperature of the other electrode to range around the jump temperature $T^J$.


%
\begin{figure*}[t]
\centering
\includegraphics[width=\textwidth]{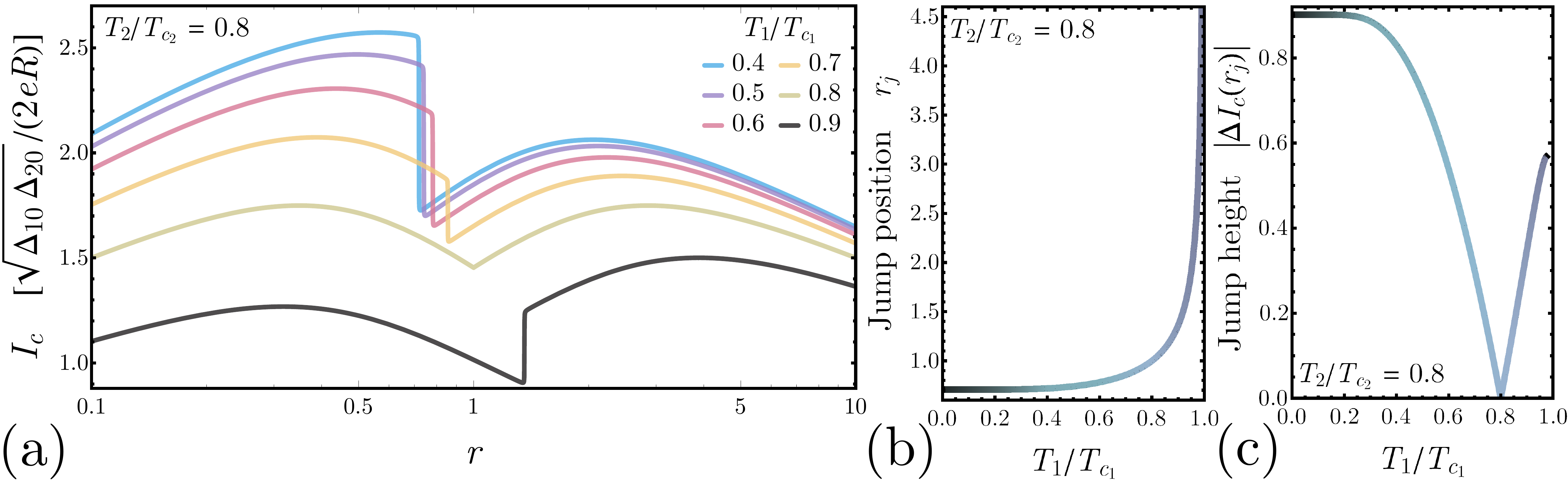}
\caption{(a), Critical current, in units of $\sqrt{\Delta_{10}\Delta_{20}}/(2eR)$, as a function of the asymmetry parameter $r$ at a few values of the normalized temperature $T_1/T_{c_1}$ and $T_2/T_{c_2}=0.8$. (b) and (c), Position and height of the critical current jump as a function of $T_1/T_{c_1}$ for $T_2/T_{c_2}=0.8$ }
\label{Fig05}
\end{figure*}

As discussed so far, the critical current strongly depends on the asymmetry parameter $r$. In this regard, in Fig.~\ref{Fig05}(a) we illustrates the behavior of the critical current $I_c$, in units of $\sqrt{\Delta_{10}\Delta_{20}}/(2eR)$, as a function of $r$, at a few values of the normalized temperature $T_1/T_{c_1}$ and $T_2/T_{c_2}=0.8$. We observe that also these curves may show a jump, except for the curve at $T_1/T_{c_1}=T_2/T_{c_2}$. In the latter case, $I_c$ shows a cusp in $r=1$, since its slope suddenly changes from negative to positive around $r=1$, and it is symmetric, in a semi-log plot, with respect to this point.
The position $r_j$ of the jump of $I_c$ changes with the temperature $T_1/T_{c_1}$ and can be estimated through Eq.~\eqref{singularities}. In Fig.~\ref{Fig05}(b), we display the jump position $r_j$ as a function of $T_1/T_{c_1}$, for $T_2/T_{c_2}=0.8$. Additionally, the height of the $I_c$ jumps, $|\Delta I_c(r_j)|$, as a function of $T_1/T_{c_1}$ is shown in Fig.~\ref{Fig05}(c) for $T_2/T_{c_2}=0.8$. We observe that $\Delta I_c(r_j)$ has a plateau at low $T_1$'s and it decreases by increasing $T_1$, up to vanish at $T_1=T_2$, whereupon it raises again.

\begin{figure*}[t]
\centering
\includegraphics[width=\textwidth]{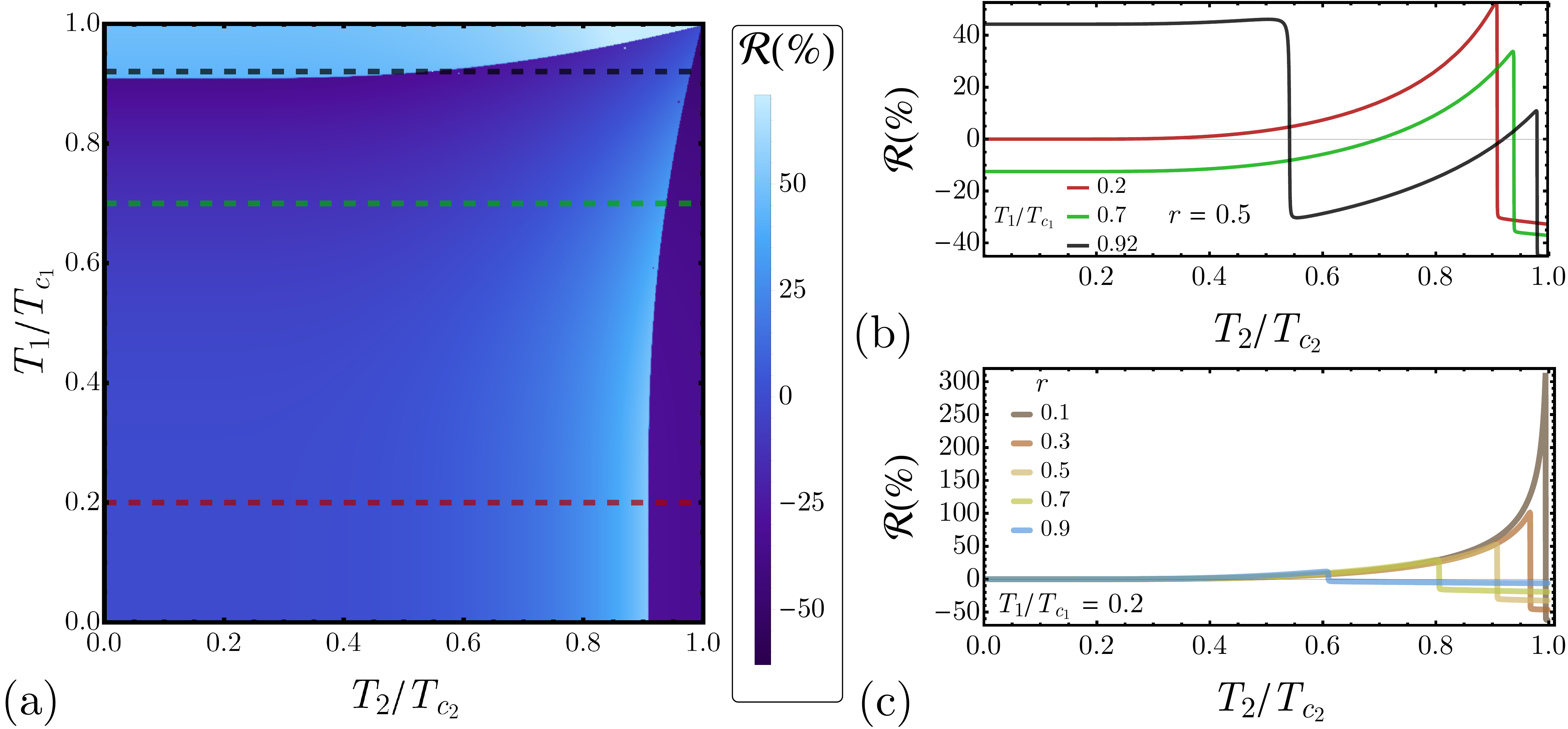}
\caption{(a) Parameter $\mathcal{R}$, see Eq.~\eqref{rectparam}, as a function of $T_1/T_{c_1}$ and $T_2/T_{c_2}$, for $r=0.5$. (b), Profiles of $\mathcal{R}$ vs $T_2/T_{c_2}$, for $T_1/T_{c_1}=\{0.2,0.7,0.92\}$ and $r=0.5$, corresponding to the colored dashed lines in (a). (c), Parameter, $\mathcal{R}$, see Eq.~\eqref{rectparam}, as a function of $T_1/T_{c_1}$, at a few values of $r$ and $T_2/T_{c_2}=0.2$. }
\label{Fig06}
\end{figure*}

Finally, with the aim to quantify the asymmetry of the critical current with respect to the switch of the temperatures $T_1$ and $T_2$, keeping fixed the structural asymmetry $r$, we define the temperature-switching asymmetry parameter $\mathcal{R}$,
\begin{equation}\label{rectparam}
\mathcal{R}(\%)=\frac{I_{c}(T_1,T_2)-I_{c}(T_2,T_1)}{I_{c}(T_2,T_1)}\times100 .
\end{equation}
This parameter synthetically describes how the structural asymmetry $r$ induces a strong asymmetrical behavior on the non-dissipative branch represented by an asymmetry of the critical current with the exchange of the temperatures of the superconducting leads.
In the density plot shown in Fig.~\ref{Fig06}(a) we display the behavior of $\mathcal{R}$ as a function of $T_1/T_{c_1}$ and $T_2/T_{c_2}$, for $r=0.5$. We observe that also $\mathcal{R}$ shows discontinuities, just in correspondence of the $I_c$ jumps previously discussed in Fig.~\ref{Fig02}. Furthermore, the sign of $\mathcal{R}$ switches in correspondence of a jump. If $\left |\mathcal{R} \right |$ is maximum, it means that the variation of $I_c$ by switching the temperatures is maximal too. Conversely, if $\mathcal{R}=0$ the critical current is symmetric with respect to a temperature switch, although the system is intrinsically asymmetric, since $r\neq1$. Three selected profiles of $\mathcal{R}$ as a function of $T_2/T_{c_2}$ for different $T_1/T_{c_1}$'s are shown as well in Fig.~\ref{Fig06}(b). The situations plotted in this figure correspond to the colored dashed lines in Fig.~\ref{Fig06}(a). For $T_1<T_1^{\text{th}}$, by varying $T_2/T_{c_2}$ we note that $\mathcal{R}$ undergoes to only one jump at a temperature $T_2>T_2^{\text{th}}$, see curves at $T_1/T_c=0.2$ and $0.7$ in Fig.~\ref{Fig06}(b). In these cases, $\mathcal{R}$ monotonically increases before the jump, whereas it becomes negative and monotonically decreases after the jump. Moreover, the height of these jump reduces by increasing $T_1$. Conversely, at a temperature $T_1>T_1^{\text{th}}$, we observe two jumps in $\mathcal{R}$, see the curve at $T_1/T_c=0.92$ in Fig.~\ref{Fig06}(b), since both $I_c(T_1,T_2)$ and $I_c(T_2,T_1)$ behaves discontinuously at some values of $T_2$. Also in this case $\mathcal{R}$ becomes negative after a jump.

The behavior of $\mathcal{R}$ as a function of $T_2/T_{c_2}$ at a few values of the asymmetric parameter $r<1$ is shown in Fig.~\ref{Fig06}(c), at $T_1/T_c=0.2$. We note that the lower the value of $r$, the higher are both the temperature at which $\mathcal{R}$ changes abruptly and the height of its jump. Conversely, in the symmetric case, $r=1$, the critical current is symmetric in the temperatures switch, namely, $I_{c}(T_1,T_2)=I_{c}(T_2,T_1)$, so that $\mathcal{R}=0$ $\forall T_1,T_2$.

\section{Linear response approximation}
\label{Sec02}\vskip-0.2cm

In this section we analyze the variation of the critical current for small temperature differences between the two superconductors. Our aim is to quantify how small temperature differences will affect the non-dissipative regime in the presence of a structural asymmetry, $r\neq0$, in the junction. We assume that $T_1>T_2$, so that we can define $T$ and $\delta T$ such that $T_1=T+\delta T/2$ and $T_2=T-\delta T/2$, and we can investigate the linear response approximation by imposing $\delta T= T_1-T_2\ll T=(T_1+T_2)/2$.

The critical current, see Eq.~\eqref{IcT1T2Im}, depends on the lead temperatures through both the statistical factors $f_j\equiv f\left ( \varepsilon ,T_j \right )$ and the self-consistent superconducting gap $\Delta_j\equiv\Delta_j(T_j)$ (with $j=1,2$). 
The linear behaviour in $\delta T$ of the critical current $I_c=\int_{-\infty}^{+\infty} d\epsilon J_c(\epsilon)$ can be easily written as

\begin{equation}\label{Icsviluppo}
\frac{\delta I_c}{\delta T} = \int_{-\infty}^{+\infty}\!\!\!\!d\epsilon\Bigg(\sum_i \underset{\alpha_1}{\underbrace{\ 
\frac{\delta J_c(\epsilon)}{\delta f_i}\bigg|_{\Delta_j}
\frac{\partial f_i}{\partial\ \delta T}
}
}
+\underset{\alpha_2}{\underbrace{\frac{\delta J_c(\epsilon)}{\delta \Delta_i}\bigg|_{f_j}
\frac{\partial \Delta_i}{\partial\ \delta T}
}}\Bigg),
\end{equation}
where in the first term ($\alpha_1$) we consider only the temperature variation of the statistical weights $f_i$ and in the second ($\alpha_2$) the temperature variation of the gaps $\Delta_i$. 
%
Finally, the critical current can be written as 
\begin{equation}
I_c\left ( T,\delta T \right )\simeq I_c(T,0)+\alpha_1(T)\delta T+\alpha_2(T)\delta T,
\end{equation}
where
\begin{equation}\label{zerocurrent}
I_c(T,0)=\frac{1}{2eR}\left |\underset{-\infty}{\overset{\infty}{\mathop \int }}f(\varepsilon , T)\text{Im}\left [ \mathfrak F_1(\varepsilon,T ) \mathfrak F_2(\varepsilon,T ) \right] d\varepsilon \right |
\end{equation}
coincides exactly with the well known Ambegaokar-Baratoff relation~\cite{Bar82}. Therefore, the linear contribution to the critical current can be seen as a correction to the usual relation, Eq.~\eqref{zerocurrent}, due to the junction asymmetry and the temperature gradient. This contribution is determined by two different terms, $\alpha_1$ and $\alpha_2$, see Eq.~\eqref{Icsviluppo}. The former is associated to the variation of the electron distribution assuming temperature-independent gaps. 
Instead, the latter, $\alpha_2$, is computed by considering only temperature variations of the superconducting gaps included in the anomalous Green's functions $\mathfrak F_j$, see Eq.~\eqref{Green}.

%
\begin{figure*}[t!!]\label{alpha1}
\centering
\includegraphics[width=0.8\textwidth]{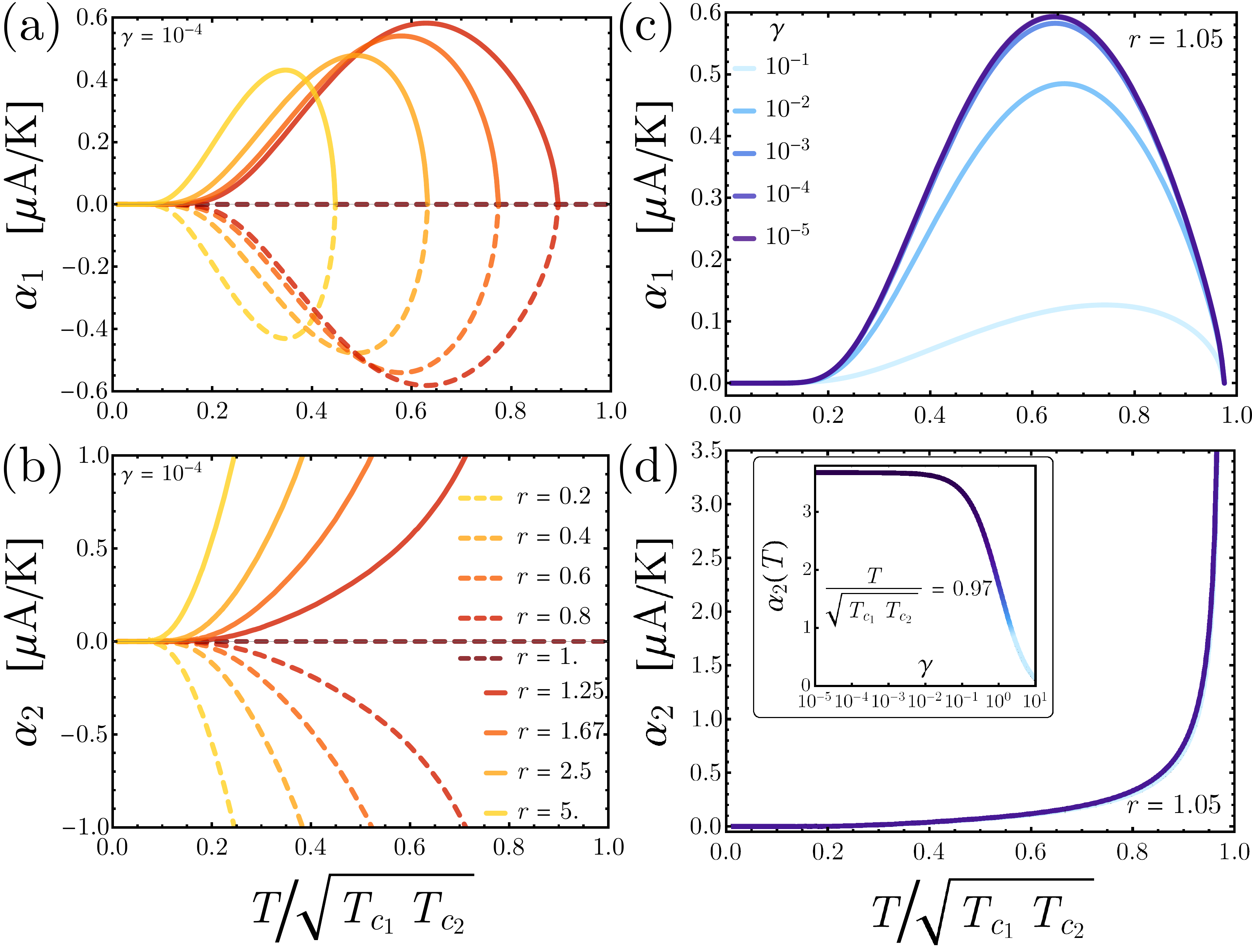}
\caption{(a) and (b), Coefficients $\alpha_1$ and $\alpha_2$, see Eqs.~\eqref{coefficente1} and~\eqref{coefficente2}, as a function of the normalized temperature $T/\sqrt{T_{c_1}T_{c_2}}$, for $\gamma=10^{-4}$ and several values of $r$. The legend in panel (b) refers to both panels. (c) and (d), Coefficients $\alpha_1$ and $\alpha_2$ as a function of the normalized temperature $T/\sqrt{T_{c_1}T_{c_2}}$ for several $\gamma$ and $r=1.05$. In the inset of panel (d): $\alpha_2$ as a function of $\gamma$ for $T/\sqrt{T_{c_1}T_{c_2}}=0.97$ and $r=1.05$. The legend in panel (d) refers to both panels.}
\label{Fig07}
\end{figure*}
%


According to the modulus in Eq.~\eqref{IcT1T2Im}, if we recast the critical current as $I_c= \left |\mathfrak{I_c} \right |$, its derivative can be written as $\frac{\partial I_c}{\partial \delta T}=\Ic\frac{\partial \mathfrak{I_c}}{\partial \delta T}$. Then
%
%
the coefficient $\alpha_1$ reads
%
\begin{equation}\label{coefficente1}
\alpha_1(T)
=\frac{\Ic}{8eRk_BT^2}\int_{-\infty }^{\infty }\!\!\!\! d\varepsilon \,\varepsilon\frac{ \text{Im}\left [\mathfrak{F}_1(\varepsilon,T) \mathfrak{F}^*_2(\varepsilon,T) \right ]}{\text{cosh}^2(\varepsilon/2k_BT)},
\end{equation}
%
where it is easy to recognise the derivative contribution of $f_i$ as taken directly from Ambegaokar-Baratoff, Eq.~\eqref{zerocurrent}.
Instead, by expanding the anomalous terms in Eq.~\eqref{IcT1T2Im} to the first order in $\delta T$, 
%
the coefficient $\alpha_2$ can be expressed as
\begin{equation}\label{coefficente2}
\alpha_2(T)=\frac{\Ic}{4eR}\underset{-\infty}{\overset{\infty}{\mathop \int }} d\varepsilon f\left ( \varepsilon ,T \right )\sum_i (-1)^{i-1}\frac{{\Delta_i}'(T) }{\Delta_i(T)}\beta _i\left( \varepsilon ,T \right ),
\end{equation}
where $\Delta_i'(T)$ is the derivative with respect to $T$ of the $i$-th superconducting gap, and $\beta_j\left( \varepsilon ,T \right )=\text{Im}\left (\mathfrak F_1\mathfrak F_2 \right )\mathfrak N^2_j-\frac{i}{2}\text{Re}\left (\mathfrak F_1\mathfrak F_2 \right )\mathfrak F^2_j$, with $\mathfrak{N}_j(\varepsilon,T ) =\left ( \varepsilon +i\Gamma_j \right )\Big/\sqrt{\left ( \varepsilon +i\Gamma_j \right )^2-\Delta^2_j \left ( T\right )}$.
We see that the gaps affect the linear coefficient $\alpha_2$ via their logarithmic derivatives $\Delta_i'(T)/\Delta_i(T)$ only. 

We note that both $\alpha_1$ and $\alpha_2$ are linear coefficients of the dissipationless regime so they can be defined only for $T\leq\text{Min}\{T_{c_1},T_{c_2}\}$. In order to efficiently represents these terms for different structural asymmetries $r$, it is convenient 
to normalize the temperature with respect to $\sqrt{T_{c_1}T_{c_2}}$. So, one can easily verify that the linear coefficients are defined only for $\frac{T}{\sqrt{T_{c_1}T_{c_2}}}\leq\text{Min}\{\sqrt{r},\frac{1}{\sqrt{r}}\}$. 

The behaviors of the coefficients $\alpha_1$ and $\alpha_2$ as a function of the normalized temperature $T/\sqrt{T_{c_1}T_{c_2}}$ for $r\in[0.2\div5]$ are shown in Figs.~\ref{Fig07}(a) and (b), respectively. 
Hereafter, we will assume a Nb ($T_{c_1}=9.2\;\text{K}$) electrode $S_1$ and we suppose to be able to set the gap of $S_2$ at will, in order to get the appropriate value of the asymmetry parameter $r$. The barrier resistance is set to $R=100\;\Omega$, which results in a junction that, for the symmetric case $r=1$, has a low-temperatures critical current approximatively of $22\;\mu\text{A}$.

First of all, we observe that both $\alpha_1$ and $\alpha_2$ vanish for $r=1$, see Figs.~\ref{Fig07}(a) and (b), respectively, namely, there is no linear contribution with the temperature gradient to the critical current in the symmetric case. Conversely, both coefficients are positive for $r>1$ and negative for $r<1$. This remark can be rationalized by observing that the critical current roughly scales according to the geometric mean of the superconducting gaps $\sqrt{\Delta_1(T_1)\Delta_2(T_2)}=\sqrt{\Delta_1(T+\delta T/2)\Delta_2(T-\delta T/2)}$. Interestingly, the $\delta T$ derivative of this quantity is positive for $r>1$ and negative for $r<1$. This shows that the sign of $\delta I_c/\delta T$ in a JJ under a small temperature gradient $\delta T$ directly reflects on the structural asymmetry $r$ in the junction.

We observe that $\alpha_1$ behaves non-monotonically, see Fig.~\ref{Fig07}(a), since, for $r<1$ it starts from zero, reaches a minimum and than it vanishes at $\frac{T}{\sqrt{T_{c_1}T_{c_2}}}=\sqrt{r}$, that is at $T=T_{c_1}$. Similarly, for $r>1$ it starts from zero, reaches a maximum and finally vanishes at $\frac{T}{\sqrt{T_{c_1}T_{c_2}}}=\frac{1}{\sqrt{r}}$, that is at $T=T_{c_2}$. For low temperatures, the behavior of $\alpha_1$ is ruled by the exponential suppression of the hyperbolic contribution for $T\to0$. Instead, for $\frac{T}{\sqrt{T_{c_1}T_{c_2}}}\to\text{Min}\{\sqrt{r},\frac{1}{\sqrt{r}}\}$, namely, for $T\to\text{Min}\{T_{c_1},T_{c_2}\}$, the product $\sqrt{\Delta_1(T)\Delta_2(T)}$ vanishes, so that $\alpha_1$ goes to zero according to the BCS temperature dependences of $\Delta_1(T)$ or $\Delta_2(T)$. Moreover, we observe that the maximum value of $\left | \alpha_1 \right |$ increases if $r\to1$. This apparently odd result is consistent with the fact that when $T_1\approx T_2$ the critical current is not-analytical in the asymmetry parameter $r$, as implied by the cusp shown in Fig.~\ref{Fig05}(a) for $r=1$ and $T_1=T_2$.

Conversely, $\alpha_2$ behaves monotonically, see Fig.~\ref{Fig07}(b). Specifically, it rapidly vanishes at $T\to0$ and diverges at $\frac{T}{\sqrt{T_{c_1}T_{c_2}}}\to\text{Min}\{\sqrt{r},\frac{1}{\sqrt{r}}\}$. The low-temperatures behavior of $\alpha_2$ is mainly governed by the gap logarithmic derivatives, being the superconducting gap roughly constant at $T\lesssim T_{c_j}/4$ so that ${\Delta_j}'(T) \to0$ at $T\to0$. Instead, for $T\to \text{Min}\{T_{c_1},T_{c_2}\}$, although $\Delta_j(T)\to0$, we observe that the logarithmic derivative diverges making $\alpha_2$ also diverging. 

Interestingly, we observe that the coefficients $\alpha_1$ and $\alpha_2$ behave quite differently by varying the Dynes parameter $\gamma$, as it is clearly shown in Figs.~\ref{Fig07}(c) and (d) for a few values of $\gamma\in[10^{-5}\div10^{-1}]$, and $r=1.05$. We observe that $\alpha_1$ is strongly affected by $\gamma$, since it significantly reduces by increasing $\gamma$, up to become even five times lower passing from $\gamma=10^ {-5}$ to $\gamma=10^ {-1}$, see Fig.~\ref{Fig07}(c). Conversely, the coefficient $\alpha_2$ is practically independent of $\gamma$, as it is shown in Fig.~\ref{Fig07}(d). Interestingly, we observe that to appreciate concrete variations in $\alpha_2$ we should consider quite higher, unrealistic values of $\gamma$, see the curve shown in the inset in Fig.~\ref{Fig07}(d) obtained at $\frac{T}{\sqrt{T_{c_1}T_{c_2}}}=0.97$.

\section{Discussion: a possible application for single-photon sensing}
\label{Sec03}\vskip-0.2cm

\begin{figure}[t!!]\label{ThermalNoise}
\centering
\includegraphics[width=\columnwidth]{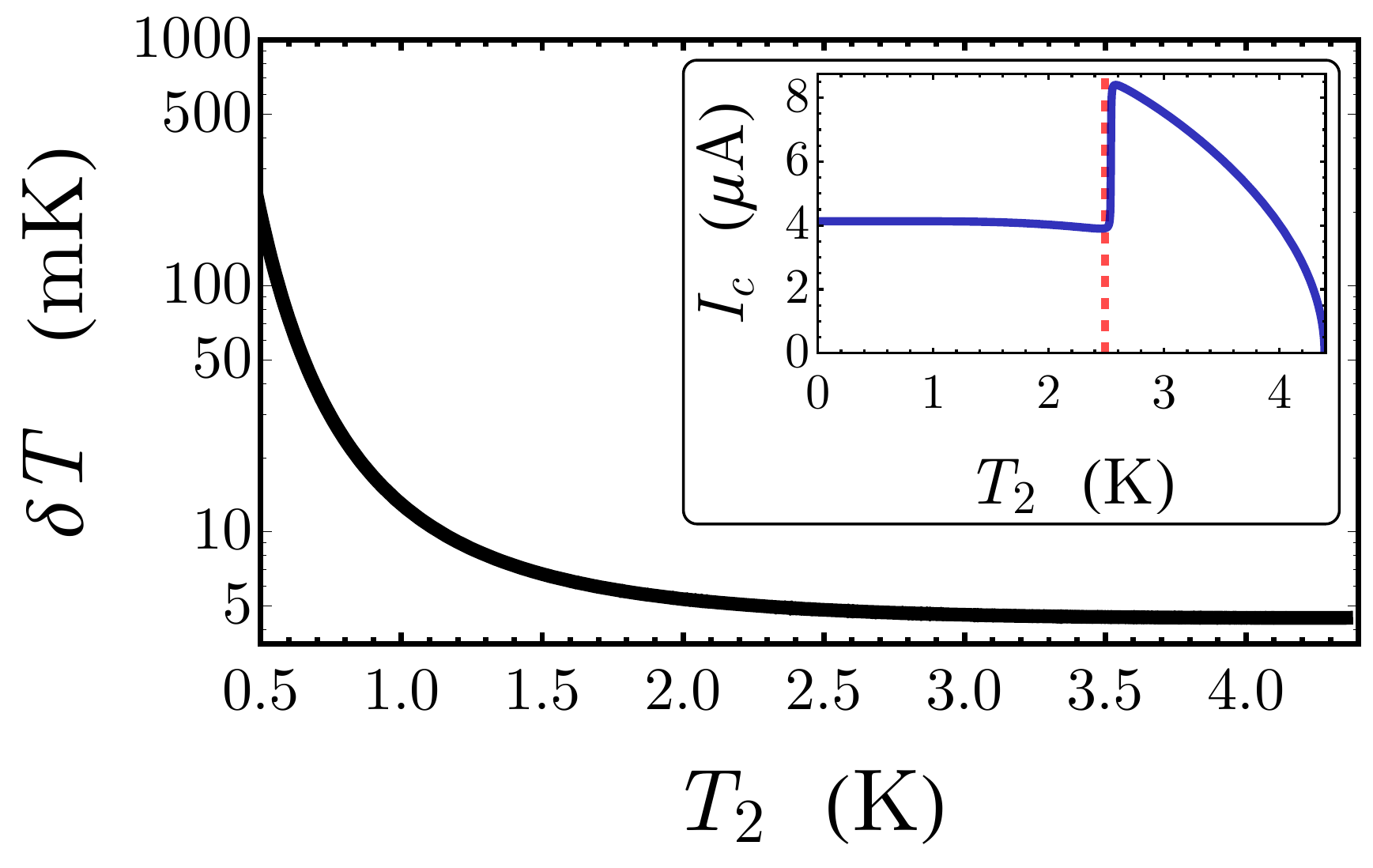}
\caption{Thermodynamic temperature fluctuations $\delta T$ as a function of the temperature $T_2$ of a Ta electrode with volume $V_2=0.01\;\mu\text{m}^3$, $T_{c_2}=4.4\;\text{K}$, and $N_{F,2}=10^{47}\;\text{J}^{-1}\text{m}^{-3}$. In the inset: critical current, $I_c$, as a function of $T_2$, at $T_1/T_{c_1}=0.93$, for a Nb/Ta junction with $R=100\;\Omega$.}
\label{Fig08}
\end{figure}

The physical effect described so far could promptly find an application in several contexts. For instance, this device can be used as the Josephson-counterpart of a thermal current rectifier, which exploits the $I_c$ drop upon temperature bias reversal. Interestingly, several examples of thermal rectifiers, namely, structures allowing high heat conduction in one direction but suppressing thermal transport upon temperatures switch, based on Josephson junctions~\cite{Mar13,GiaBer13,ForMar15,Mar15}, phononic devices~\cite{Cha06,Kob09,Tia12}, and quantum dot~\cite{Sch08}, were also recently conceived.

Alternatively, a non-dissipative single-photon detector~\cite{Wei08,Vou10,Che11,Ber13,Sol17,Wal17,Vir18} based on a temperature-biased asymmetric Josephson tunnel junction might be conceived. The development of superconducting sensors is receiving a growing interest recently, since the use of superconducting devices working at cryogenic temperatures has some advantages. To name just a few, they assure a significant suppression of the heat leakage~\cite{Kar00,Wei08,Cha18,Hei18} and a vanishingly small Johnson noise~\cite{Gia08}, since operating in the non-dissipative regime.

The setup we are proposing resembles a superconducting tunnel junction (STJ) detector where a tunnel Josephson junction is exploited in the dissipative regime~\cite{Pea96,Pea97,Pea98}. Conversely, in our proposal we operate the tunnel junction in the dissipationless regime without involving any quasiparticle charge current~\footnote{In the presence of the thermal gradient there will be a quasiparticle heat current diffusion between the two superconductors but, due to particle-hole symmetry, there is not any thermoelectrical current associated to it.}.

This single-photon detector implementation is worth to be discussed in more detail. In the proposed device concept, the measurable abrupt increase of the critical current, as determined by the enhancement of the temperature of the absorbing superconducting layer, could be exploited to detect radiation. In such a setup, the photon is supposed to be absorbed in an electrode with a small volume (i.e., with a small heat capacitance), for instance, the electrode $S_2$, to allow its temperature to effortlessly change due to a small energy absorption. The other electrode, i.e., $S_1$, is instead supposed to have a large volume and it is endowed with a heating probe continuously injecting heat, in order to maintain its temperature $T_1$ as constant as possible. At the same time, the system is assumed in good thermal contact with a phonon bath. 

The temperature of the electrode $S_2$ is initially kept close to the threshold value $T_2^J$. Due to the photon absorption and the resulting temperature increase, the critical current can jump. Then, in such a detector, the choice of the working temperature, $T_2(0)$, is an essential point. Especially, $T_2(0)$ must be chosen close enough to $T_2^J$, so that the incoming photon can induce the transition.
At the same time, if $T_2(0)$ is too close to $T_2^J$, unavoidable thermal fluctuations in the superconductor could trigger a faulty detector reaction.
For this reason, the analysis of the thermal fluctuations is of crucial importance to estimate the detector feasibility.
Indeed, we need first of all the temperature separation $\Delta T_2=T_2^J-T_2(0)$ to be much larger than possible temperature fluctuations, in order to reduce dark counts to a minimum. At the same time, reducing the separation $\Delta T_2$ increases the sensitivity to low energy photons. The thermodynamic temperature fluctuations can be estimated as~\cite{Wal17,Bra18}
\begin{equation}
\delta T=\sqrt{\frac{k_BT^2}{C_j(T)}}.
\end{equation}
Here, $C_j(T)=T \,\partial \mathcal{S}_j/\partial T$ is the electronic heat capacity of the superconductor $S_j$, where $\mathcal{S}_j(T)$ is its electronic entropy and it is given by~\cite{Rab08,Sol16}
\begin{equation}
\mathcal{S}_j(T)=-4k_BV_jN_{F,j}\int_{-\infty}^{\infty} f(\varepsilon,T) \ln[ f(\varepsilon,T)] \mathcal{N}_j(\varepsilon,T) d\varepsilon.
\label{Entropy}
\end{equation}
Here $V_j$ is the volume, $N_{F,j}$ is the quasiparticle density of states at the Fermi energy, $f ( E ,T )$ is the Fermi distribution function, and $\mathcal{N}_j\left ( \varepsilon ,T \right )=\left | \text{Re}\left [ \frac{ \varepsilon +i\Gamma_j}{\sqrt{(\varepsilon +i\Gamma_j) ^2-\Delta _j\left ( T \right )^2}} \right ] \right |$ is the smeared BCS density of states of $S_j$.
Since $\delta T\propto V_j^{-\frac{1}{2}}$, we observe that the lower the electrode volume, the higher the thermal fluctuations. So, the detector design should carefully determine the volume of the absorber, since a smaller volume may be beneficial for sensitivity at low energies, but also potentially detrimental due to the increase of thermodynamic fluctuations.

Hereafter we refer to a device configuration previously taken as an example, namely, a junction with electrodes $S_1$ and $S_2$ respectively made by Nb ($T_{c_1} = 9.2\;\text{K}$) and Ta ($T_{c_2} = 4.4\;\text{K}$), so that $r\approx2$. This material selection has already demonstrated a high quantum efficiency in absorbing photons from IR to UV frequencies~\cite{Pea98}. In the inset of Fig.~\ref{Fig08}, we show the critical current of a Nb/Ta junction with $R=100\;\Omega$ as a function of $T_2$. In this case, the critical current jumps at $T_2\simeq2.54\;\text{K}$. 
The behavior of $\delta T$ as a function of the temperature of the Ta electrode with $V_2=0.01\;\mu\text{m}^3$, $T_{c_2}=4.4\;\text{K}$, and $N_{F,2}=10^{47}\;\text{J}^{-1}\text{m}^{-3}$, is shown in Fig.~\ref{Fig08}. We observe that in this case the fluctuations are vanishingly small in a large range of temperatures. For instance, by assuming to work at $2.5\;\text{K}$ one obtains $\delta T\simeq5\;\text{mK}$. This means that a working temperature far, for instance, just $\Delta T_2=50\;\text{mK}$ from the threshold value could safely prevent an untrustworthy absorber temperature readout. 
The red dashed line in the inset of Fig.~\ref{Fig08} indicates the working temperature obtained by choosing $\Delta T_2=50\;\text{mK}$.

We observe that in the single-photon detection mode, the proposed detector is characterized by a ``dead time'' in which it cannot be used to reveal a following incident photon. After an absorption, the temperature $T_2$ increases reaching a maximum during a jitter time and then, due to the thermal contact with the phonon bath, the electrode $S_2$ recovers its initial steady temperature. However, once a transition induced by a photon with enough energy has occurred, a further photon-induced temperature increase would not induce another $I_c$ jump, unless the system has already switched back to its idle state. During the thermal evolution following a photon absorption, the condition $T_2=T_2^J$ at which $I_c$ jumps is satisfied twice. The distance in time between these subsequent photon-induced $I_c$ jumps can be used to define the dead time of the device. Since the maximum temperature reached by $S_2$ depends on the absorbed energy, the photon frequency could be directly inferred from this dead time, that can be reduced by a device and fabrication optimization. In fact, since the thermalization time can be estimated as $\tau_{\text{th}}=C_j/G$~\cite{Gua18} (with $G$ being the total thermal conductance of the JJ), the energy excess due to the photon absorption could be released more quickly by allowing the superconductor $S_2$ to be strongly coupled to the thermal phononic bath. The possibility to work at temperatures of the order of $T_c/2$ guarantees a good {\it{e-ph}} coupling, and then a quite short dead time in comparison with other detectors working at $T\ll T_c$.
Furthermore, in the case of monochromatic radiation, our device shows unique photon-number-resolving detection capabilities, since the dead time directly depends on the absorbed energy.

\begin{figure}[t!!]\label{ResolvingPower}
\centering
\includegraphics[width=\columnwidth]{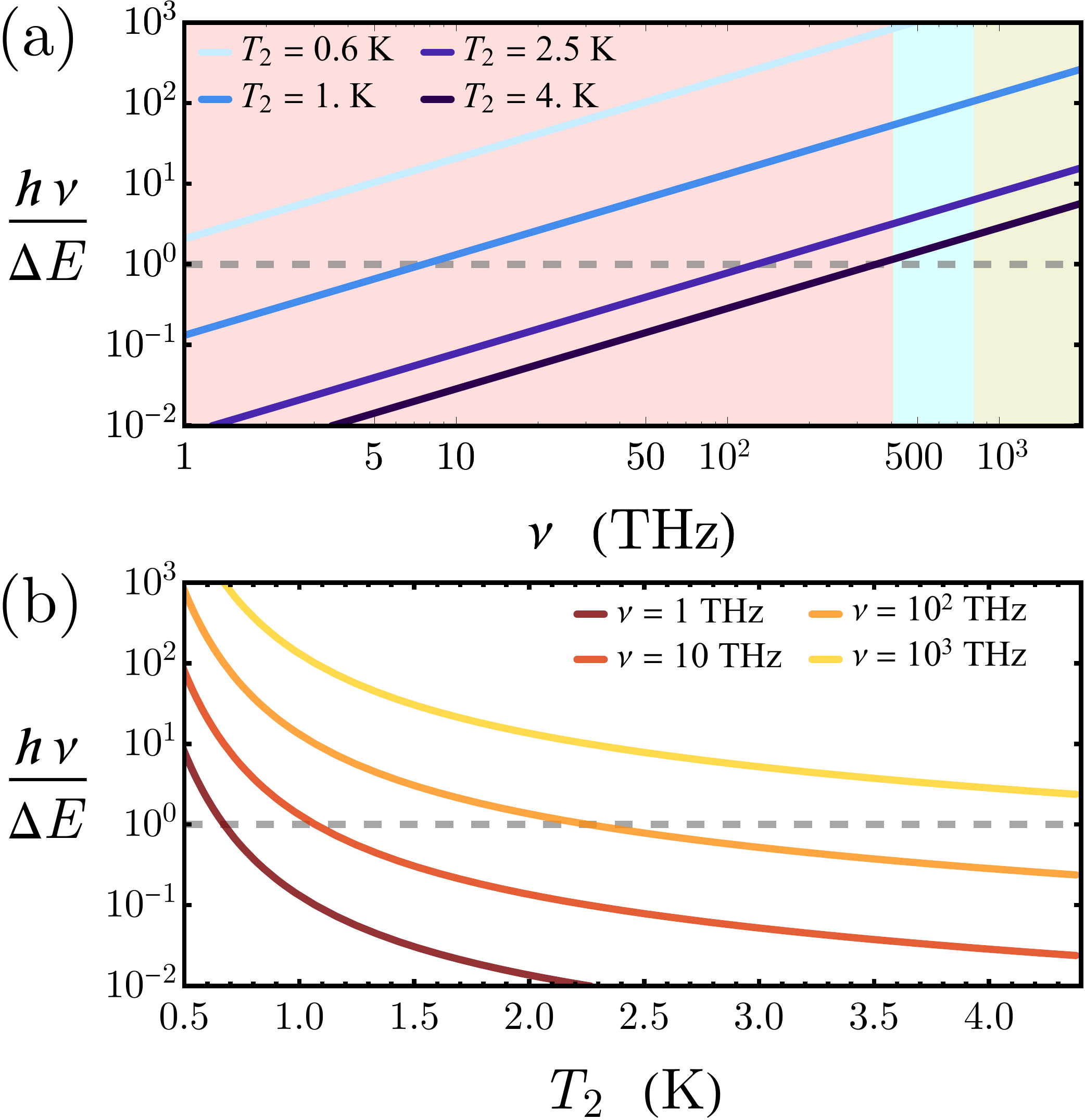}
\caption{(a) Resolving power as a function of the photon frequency at a few temperatures. The shaded regions indicate the frequency ranges corresponding to IR (red), visible (cyan), and UV (green) light spectrum. (b) Resolving power as a function of the temperature at a few values of the photon frequency. The values of other parameters are: $V_2=0.01\;\mu\text{m}^3$, $T_{c_2}=4.4\;\text{K}$, and $N_{F,2}=10^{47}\;\text{J}^{-1}\text{m}^{-3}$.}
\label{Fig09}
\end{figure}

The possibility of distinguishing photons with different frequencies would allow us to use the device as a calorimeter. To estimate the performance of a calorimeter the relevant figure of merit is the resolving power, which is calculated in the idle state in the absence photonic excitation, and reads~\cite{Vou10,Vir18}
\begin{equation}
\frac{h\nu}{\Delta E}=\frac{h\nu}{4\sqrt{2 \ln 2}\sqrt{k_BT^2C(T)}},
\end{equation}
where $T$ is the steady temperature of the absorber, $\nu$ is the photon frequency, and $\Delta E$ is the intrinsic energy resolution of full width at half maximum for a calorimeter with a white-noise spectrum~\cite{Gia06,Vou10}.
Fig.~\ref{Fig09}(a) shows the resolving power as a function of the photon frequency, $\nu$, at a few temperatures of the Ta electrode with volume $V_2=0.01\;\mu\text{m}^3$. We observe that the resolving power obviously increases linearly with the photon frequency. The horizontal dashed line indicates unitary resolving power. We note that at $0.6\;\text{K}$ a resolving power exceeding one results in the whole range of frequencies shown in Fig.~\ref{Fig09}(a) (infrared to UV light spectrum). Instead, for $T_2=2.5\;\text{K}$, namely, the working temperature previously discussed, see the inset of Fig.~\ref{Fig08}, we obtain $h\nu/\Delta E>1$ only at frequencies above $100\;\text{THz}$. This means that a Nb/Ta-based detector, with the chosen detection volume $V_2$, residing at a temperature $T_2=2.5\;\text{K}$ could properly work as a calorimeter for frequencies $\nu\gtrsim100\;\text{THz}$. 

The temperature dependence of the resolving power at a few values of the photon frequency is displayed in Fig.~\ref{Fig09}(b). We note that the resolving power monotonically reduces with increasing the temperature, and that the higher $\nu$, the larger the range of temperatures giving $h\nu/\Delta E>1$. 

We estimate the sensitivity of the device by assuming some thermal response to the photon absorption. However, we need to discuss how critical current measurements can be done without affecting previous conclusions. 
Reading of the photon-induced $I_c$ variation could be performed by conventional techniques, for instance, via a Josephson sensor~\cite{Gia08} based on the modifications of the kinetic inductance, $L_k\propto1/I_c$~\cite{Bar82,Lik86}, of the junction working in the dissipationless regime and inductively coupled to a superconducting quantum interference device (SQUID). Alternatively, the variation of the Josephson kinetic inductance of the junction can be performed dispersively through an LC tank circuit inductively coupled to the JJ~\cite{Gov14,Gov16}. As a matter of fact, in this readout scheme the modifications of the Josephson inductance can be measured through a shift, or a broadening, of the circuit's transmission or reflection resonance~\cite{Day03}.
Detectors based on a dispersive detection have a huge potential in fast detection and quantum limited energy-resolution~\cite{Gov16}. Those platforms combined with the dissipationless configuration of our tunnel junction could promise minimal low-noise performances with reduced dark-counts and, consequently, high energy sensitivity.
In this dispersive configuration, one can also deploy multiplexing capabilities, paving the way to the real-time control of more single-photon sensors, making attractive platform for astrophysical applications. 

Finally, we observe that actually our detection proposal shows up some similarities with other single-photon sensors based on the critical current change due to the photon absorption in a proximized nanowire~\cite{Vou10,Gov16,Vir18,Sol17}. Anyway, there are also several qualifying differences. Firstly, in our detection scheme the absorbing element is a superconducting lead of an asymmetric JJ and the phenomenon exploited for the detection is the anomalous steep variation by changing the temperature of the critical current, where its temperature variation is smoother in proximized sensors. Therefore, the strength of our device resides in a strong sensitivity due to the steeplike response of $I_c$ to a photon-induced heating. Moreover, the fact that the detection is not performed in extremely low temperatures regimes, could be advantageous for achieving a fast thermal response due to a better {\it{e-ph}} coupling, which results in a shorter dead time of the detector. Markedly, we think that our detector represents an interesting combination between different types of superconducting single-photon and calorimetric devices. In particular, it has the potential sensitivity of STJ systems, however without being affected by the Johnson-Nyquist noise, due to the dissipationless working regime. Besides, the proposed detector has potentially the energy sensitivity of proximity-based detectors, with reduced dead time at parity of photon energy, due to higher operating temperatures. Finally, it is characterized by a fast thermal response, due to the energy absorption with a short timing jitter, similarly to transition-edge sensors. In conclusion, we wish to stress that the presented analysis has not been specifically optimized in performance, but it was simply done on the base of realistic and feasible parameters. We will deserve a more detailed analysis of both the detector design and its performance figures of merit in a forthcoming paper~\cite{Gua19}.

Before concluding, we wish to remark that our prediction of a jump in the critical current, in the presence of both an asymmetry of the junction and a temperature bias, is purely based on a conventional BCS mechanism, i.e., gaps matching. This means that for all those experiments where jumps in the critical current are indeed discussed as a smoking-gun proof of more elaborate mechanisms, such as, for instance, topological transitions~\cite{Marra16,Tiira17,Cayao17}, one need to deserve extra care, in order to be sure that a structural asymmetry, in the presence of an uncontrolled thermal gradient evolution, could eventually provide a simpler explanation.

\section{Conclusions}
\label{Sec04}\vskip-0.2cm

In conclusion we discuss in this paper the behavior of the critical current, $I_c$, of a Josephson tunnel junction formed by different superconductors. We analyze in detail the behavior of $I_c$ by changing both the temperatures of the electrodes and the ratio, $r$, between the critical temperatures of the superconductors. We observe that the critical current is asymmetric in the temperatures switch and that it shows a steeplike behavior at specific temperatures, namely, at the temperatures at which the BCS superconducting gaps coincides. Specifically, in these conditions the critical current of an asymmetric junction, i.e., $r\neq1$, suddenly jumps. We observe also an unexpected behavior, since, for $r>1$, by enhancing the temperature the critical current in correspondence of a jump increases. 

Studying the height of the $I_c$ jump, we observe a non-monotonic behavior, according to which we found that an optimal $r$ value, giving a maximum increase of the critical current upon temperature variations, exists. We also discuss how Dynes parameters in the superconductors affect the sharpness of the $I_c$ transition. Finally we discuss in detail the behavior of the critical current for a small thermal gradient along the junction as a function of the average temperature and the Dynes parameters.

The peculiar temperature-dependence of the critical current of an asymmetric Josephson junction can be relevant to conceive intriguing applications. For instance, the step-like variation with the temperature of the critical current will allow us to design a single-photon threshold detector in which the absorption of a photon produces a temperature enhancement, that can correspond to a measurable critical current variation. This system operating in the non-dissipative branch is likely to provide very-high energy sensitivity. The conceived device is inherently energy resolving, and can be also engineered to determine the photon number in the case of a monochromatic source of light. We briefly discussed the essential figures of merit of this type of detector, which deserve further investigation and a more careful design optimization, in order to address better its intrinsic potential. 

\begin{acknowledgments}
This research was supported in part by the National Science Foundation under Grant No. NSF PHY17-48958. C.G., A.B., and F.G. acknowledge the European Research Council under the European Union’s Seventh Framework Program (FP7/2007-2013)/ERC Grant agreement No. 615187-COMANCHE and the Tuscany Region under the PAR FAS 2007-2013, FAR-FAS 2014 call, project SCIADRO, for financial support.
P.S. and A.B. have received funding from the European Union FP7/2007-2013 under REA Grant agreement No. 630925 -- COHEAT. 
A.B. acknowledges the CNR-CONICET cooperation programme ``Energy conversion in quantum nanoscale hybrid devices'' and the Royal Society though the International Exchanges between the UK and Italy (grant IES R3 170054).
\end{acknowledgments}


%

\end{document}